# Analytical Realization of Finite-Size Scaling for Anderson Localization: Is There a Transition in the 2D Case?

## I. M. Suslov

*Kapitza Institute for Physical Problems, Russian Academy of Sciences, Moscow, 117337 Russia*
*e-mail: suslov@kapitza.ras.ru*

**Abstract**—Roughly half the numerical investigations of the Anderson transition are based on consideration of an associated quasi-1D system and postulation of one-parameter scaling for the minimal Lyapunov exponent. If this algorithm is taken seriously, it leads to unambiguous prediction of the 2D phase transition. The transition is of the Kosterlitz–Thouless type and occurs between exponential and power law localization (Pichard and Sarma, 1981). This conclusion does not contradict numerical results if raw data are considered. As for interpretation of these data in terms of one-parameter scaling, this is inadmissible: the minimal Lyapunov exponent does not obey any scaling. A scaling relation is valid not for a minimal, but for some effective Lyapunov exponent whose dependence on the parameters is determined by the scaling itself. If finite-size scaling is based on the effective Lyapunov exponent, the existence of the 2D transition becomes not definite, but still rather probable. Interpretation of the results in terms of the Gell-Mann–Low equation is also given.

## 1. INTRODUCTION

The one-parameter scaling hypothesis [1] leads to a conclusion that there is no Anderson transition in two dimensions. This statement has produced a breakthrough in the physics of disordered systems and led to development of the concept of weak localization with numerous experimental manifestations [2]. The recent discovery of the 2D metal-insulator transition [3–6] threatens to undermine the basic concepts of the theory. It is still unclear whether this transition can exist for a purely potential scattering or should it be related to different complications, such as interaction, spin-orbit effects, etc. It will be shown below that the first possibility is rather probable and does not suggest substantial revision in the weak localization region [1].

Initially, the present investigation was motivated by analysis of the methodical aspects of finite-size scaling [7], which is a basic concept of all recent numerical studies of the Anderson transition [8–23]. There is the problem that numerical results have a tendency to contradict all other information on the critical behavior [24]. Practically all theoretical and experimental investigations agree with the result of the Vollhardt and Wölfle self-consistent theory [25, 26]

$$\nu = \begin{cases} 1/(d-2), & 2 < d < 4 \\ 1/2, & d > 4, \end{cases} \qquad (1)$$

$$s = 1, \quad 2 < d < \infty,$$

where $\nu$ and $s$ are critical exponents of the correlation length and conductivity and $d$ is dimensionality of space. Indeed, the result in (1)

(a) distinguishes values $d_{c1} = 2$ and $d_{c2} = 4$ as the lower and the upper critical dimensions;[1]

(b) agrees with the result[2] for $d = 2 + \epsilon$ [33]

$$\nu = \frac{1}{\epsilon} + 0 \cdot \epsilon^0 + 0 \cdot \epsilon^1 + O(\epsilon^2); \qquad (2)$$

(c) agrees with the results $\nu = 1/2$ [35, 36] and $s = 1$ [37] for $d = \infty$;

(d) satisfies the scaling relation $s = (d-2)\nu$ for $d < d_{c2}$ [1];

(e) gives critical exponents independent of $d$ for $d > d_{c2}$, as is usual for mean field theory;

---

[1] The first is a consequence of one-parameter scaling [1], and the second can be seen from different points [27, 28], the main of which is renormalizability. The theory of disordered systems is mathematically equivalent to the $\varphi^4$ field theory with a "wrong" sign of interaction [29–31]. The latter is renormalizable for $d \le 4$ and nonrenormalizable for $d > 4$ [32]. For $d \le 4$, all of the physics is determined by small momenta or large distances, in accordance with the expected scale invariance. For $d > 4$, the atomic scale cannot be excluded from the results and no scale invariance is possible.

[2] According to Wegner [34], the term of order $\epsilon^2$ in (2) is finite, large, and negative. However, this result was derived for the zero-component $\sigma$-model, whose correspondence with the initial disordered system is approximate and valid for small $\epsilon$; therefore a difference can arise in a definite order in $\epsilon$.

(f) agrees with experimental results for $d = 3$, $s \approx 1$, $\nu \approx 1$ [38, 39].[3] As for numerical results, they can be summarized by the empirical formula $\nu \approx 0.8/(d-2) + 0.5$ [17], which has evident fundamental defects [24].

The finite-size scaling approach is based on the philosophy that any dimensionless quantity $A$ related to a system spatially restricted on a scale $L$ is a function of a ratio $L/\xi$,

$$A = F(L/\xi), \qquad (3)$$

where $\xi$ is the correlation length. To justify Eq. (3), let us assume that the dependence of $A$ on the parameters can be expressed as its dependence on characteristic length scales $L$, $\xi$, $l_1$, $l_2$, …. Taking $\xi$ as a unit scale, we can write

$$A = F(L/\xi, l_1/\xi, l_2/\xi, \ldots). \qquad (4)$$

Near the critical point, the correlation length $\xi$ is large in comparison with microscopic scales $l_1$, $l_2$, … and substitution $l_1/\xi = l_2/\xi = \ldots = 0$ reduces (4) to (3). This derivation is based on assumption that limiting transition $l_i/\xi \longrightarrow 0$ is not singular and the right-hand side of Eq. (4) does not become zero or infinity. Unfortunately, there is no simple way to establish when such an assumption is true.[4] When Eq. (3) is valid, it makes it possible to investigate the dependence of $\xi$ on parameters. If something is wrong with Eq. (3), it leads to erroneous conclusions.

Below we present an analytical realization of the commonly used variant of finite-size scaling based on the concept of the minimal Lyapunov exponent. Our approach is based on an investigation of the second moments for a solution of the Cauchy problem for the Schrödinger equation (Section 3), and in this respect it is close to [42, 43]. Nevertheless, justification of the approach (Section 2) and interpretation of the results (Section 3) are essentially different, and in fact we disagree with most of the statements made in [42, 43].

Briefly, our results consist in the following. If the concept of the minimal Lyapunov exponent is taken seriously, it leads to unambiguous prediction of the 2D phase transition (Section 3). The transition occurs between exponential and power law localization and, consequently, it is of the Kosterlitz–Thouless type [7]. This conclusion does not contradict the numerical results [8–13] if the raw data are considered (Section 4). Interpretation of these data in terms of one-parameter scaling is inadmissible: the minimal Lyapunov exponent does not obey any scaling. We argue that a scaling relation is valid not for minimal, but for some effective Lyapunov exponent whose dependence on parameters is determined by scaling itself (Section 5). After such modification, existence of the 2D transition becomes not definite, but still rather probable (Section 6). Interpretation of results in terms of the Gell-Mann–Low equation [1] is given in Section 7.

## 2. BASIC CONCEPTS

**2.1.** The concept of finite-size scaling is taken from the theory of phase transitions [44–46] and can be discussed using a ferromagnet as an example. Instead of an infinite 3D system, let us consider a system of size $L \times L \times L_z$ with $L_z \longrightarrow \infty$. Such a system is topologically one-dimensional and does not exhibit phase transition. The correlations in it are always paramagnetic and there is a finite correlation length $\xi_{1D}$. Relation of $\xi_{1D}$ to the ferromagnetic phase transition in the 3D system is expressed by the following statements. If $T > T_c$ and the 3D system is paramagnetic, then $\xi_{1D}$ obviously coincides with the correlation length $\xi$ of the 3D system when $L$ is sufficiently large:

$$\xi_{1D} \longrightarrow \xi \quad \text{for} \quad L \longrightarrow \infty. \qquad (5)$$

If $T < T_c$ and the 3D system is ferromagnetic, then the following statement is valid:

$$\frac{\xi_{1D}}{L} \longrightarrow \infty \quad \text{for} \quad L \longrightarrow \infty, \qquad (6)$$

which can be proved by contradiction. Indeed, let the ratio $c = \xi_{1D}/L$ be finite for all $L$. Let us assume $n \gg c$ and consider a system of size $L \times L \times nL$. The correlations in the length direction are paramagnetic and the average (along the cross-section) magnetic moment changes its sign many times. This situation holds for all $L$ and, in particular, for $L \longrightarrow \infty$; however, such a thermodynamic limit is topologically three-dimensional and a system should become ferromagnetic. This contradiction proves (6).

If $T = T_c$, then any behavior

$$\xi_{1D} \propto L^\alpha \quad (0 < \alpha \leq 1) \quad \text{for} \quad L \longrightarrow \infty \qquad (7)$$

is possible. Indeed, the ratio $c = \xi_{1D}/L$ is finite or tends to zero, and the above-considered system of size $L \times L \times nL$ possesses paramagnetic correlations. Nevertheless, it is not a true paramagnet, because its correlation

---

[3] These remarkable properties of result (1) arouse suspicions that it is exact [40]. In fact it can be obtained without model approximations on the basis of symmetry considerations [41].

[4] Such a possibility exists in the field-theoretical formulation of the problem. When the maximum microscopic scale $l_1$ tends to zero, the theory becomes divergent. In nonrenormalizable theories, such divergences are unavoidable and relation (3) never holds. If a theory is renormalizable, all divergencies can be absorbed in a finite number of parameters (such as mass, coupling constant, etc.) so that renormalized Green functions (and quantities that can be expressed via them) do not depend on $l_i$ and exhibit scale invariance. If quantity $A$ has no clear field-theoretical interpretation, it is difficult to establish its independence on the "bare mass," "bare coupling," etc. The latter quantities essentially depend on $l_i$ and are observable in condensed matter applications.

length $\xi \sim \xi_{1D}$ is divergent, as it should be at the critical point.

Usually, relation (7) is suggested with $\alpha = 1$, because it is the only possibility compatible with scale invariance. Indeed, if the quantities $\xi_{1D}$, $\xi$, and $L$ are related by some functional relation which does not contain any other scales, this relation assumes the form $F(\xi_{1D}/L, \xi/L) = 0$ if $L$ is taken as the unit length. Solving this relation for $\xi_{1D}/L$ we have

$$\frac{\xi_{1D}}{L} = F\left(\frac{L}{\xi}\right), \qquad (8)$$

and $\xi_{1D} = F(0)L$ at the critical point in accordance with (7) for $\alpha = 1$.

As a result, the quantity

$$g(L) = \frac{\xi_{1D}}{L} \qquad (9)$$

can be taken as a scaling variable whose dependence on $L$ is shown in Fig. 1a. It should be stressed, however, that $\xi_{1D}$ is sensible to the 3D transition independently of the existence of scale invariance. The latter is absent for space dimensions $d > 4$ in the case of a ferromagnet.

**2.2.** Application of these considerations to the localization theory is based on identification of $\xi_{1D}$ with the inverse of the minimal Lyapunov exponent $\gamma_{min}$,

$$\xi_{1D} \sim \frac{1}{\gamma_{min}}. \qquad (10)$$

The Lyapunov exponents occur in the solution of the Cauchy problem for the quasi-1D Schrödinger equation with the initial conditions on the left edge of the system. For example, the 1D Anderson model

$$\psi_{n+1} + \psi_{n-1} + V_n\psi_n = E\psi_n \qquad (11)$$

can be rewritten in the form of the recurrence relation

$$\begin{vmatrix} \psi_{n+1} \\ \psi_n \end{vmatrix} = \begin{vmatrix} E - V_n & -1 \\ 1 & 0 \end{vmatrix} \begin{vmatrix} \psi_n \\ \psi_{n-1} \end{vmatrix} \equiv T_n \begin{vmatrix} \psi_n \\ \psi_{n-1} \end{vmatrix}, \qquad (12)$$

where $T_n$ is a transfer matrix. Then the initial condition problem can be formally solved as

$$\begin{vmatrix} \psi_{n+1} \\ \psi_n \end{vmatrix} = T_n T_{n-1} \ldots T_2 T_1 \begin{vmatrix} \psi_1 \\ \psi_0 \end{vmatrix}. \qquad (13)$$

An analogous relation occurs for an arbitrary quasi-1D system if the quantity $\psi_n(r_\perp)$, depending on the transverse coordinate $r_\perp$, is considered as a vector $\psi_n$.

One can try to represent a matrix product $P_n$ in Eq. (13) as the $n$th power of a constant matrix $T$. Such a thing is possible, but only for the "Hermitian part" of $P_n$. As any matrix, $P_n$ can be represented as a product of the unitary matrix $U_n$ and the Hermitian matrix $H_n$

$$\begin{aligned} P_n &= T_n T_{n-1} \ldots T_2 T_1 \equiv U_n H_n, \\ H_n^2 &= P_n^+ P_n, \end{aligned} \qquad (14)$$

where $H_n$ has real eigenvalues and describes a systematic growth or decrease $\psi_n$, while $U_n$ has eigenvalues with the unit modulus and describes an oscillatory behavior. Representation $H_n = T^n$ is constructive, because the geometric mean of matrices

$$\begin{aligned} T &= (P_n^+ P_n)^{1/2n} \\ &= (T_1^+ T_2^+ \ldots T_n^+ T_n \ldots T_2 T_1)^{1/2n} \end{aligned} \qquad (15)$$

tends to a nonrandom limit for $n \longrightarrow \infty$ according to the Oseledec theorem [47]. If a vector of initial conditions in (13) is expanded in eigenvectors of $T$, while its eigenvalues $\lambda_s$ are written as $\exp(\gamma_s)$, then the following decomposition is valid for $\psi_n(r_\perp)$:

$$\begin{aligned} \psi_n(r_\perp) &= A_1 h_n^{(1)}(r_\perp) e^{\gamma_1 n} \\ &+ A_2 h_n^{(2)}(r_\perp) e^{\gamma_2 n} + \ldots + A_m h_n^{(m)}(r_\perp) e^{\gamma_m n}. \end{aligned} \qquad (16)$$

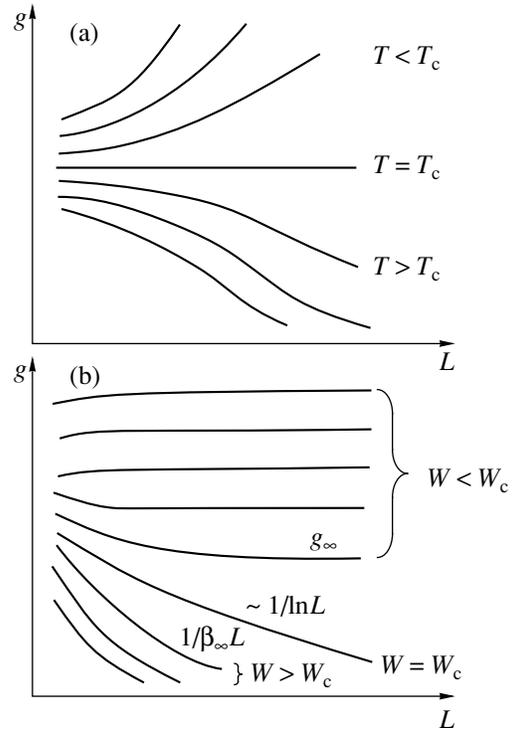

**Fig. 1.** (a) Typical dependences $g(L)$ in the case of one-parameter scaling; (b) dependences $g(L)$ according to Eq. (50).

The quantities $h_n^{(s)}(r_\perp)$ have no systematic growth in $n$, while the Lyapunov exponents $\gamma_s$ tend to constant val-ues in a large $n$ limit. Only terms with positive $\gamma_s$ are kept in Eq. (16) and they are numerated in the order of decreasing $\gamma_s$.

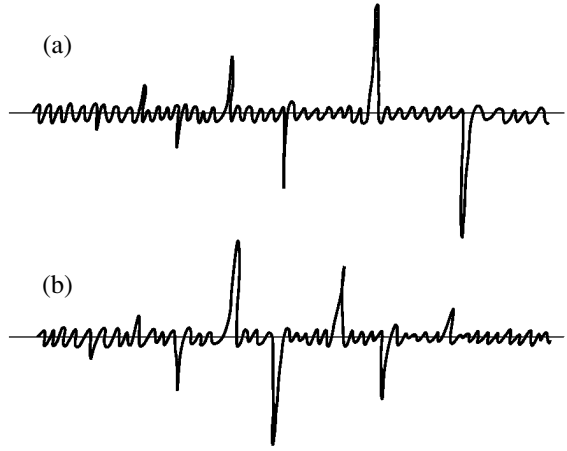

**Fig. 2.** Solution of the Cauchy problem (a) and a 1D eigenfunction constructed according to Mott (b) in the situation $a = 0$, $b > 0$.

Following to Mott [48], we can construct eigenfunctions of a quasi-1D system by matching two solutions of the type (16) increasing from two opposite edges of the system. The tails of the eigenfunction will be determined by the minimal Lyapunov exponent $\gamma_{\min} \equiv \gamma_m$ and these are grounds for relation (10).

**2.3.** Decomposition (16) is valid for nonaveraged quantity $\psi_n(r_\perp)$ and its meaning consists in distinguishing the self-averaging exponents $\gamma_s$. It will be shown in Section 3 that the mean value of $\psi_n(r_\perp)$ does not obey systematic growth,

$$\langle \psi_n(r_\perp) \rangle \sim 1, \qquad (17)$$

while decomposition of type (16) is valid for its second moment

$$\langle \psi_n^2(r_\perp) \rangle = B_1(r_\perp) e^{\beta_1 n} + B_2(r_\perp) e^{\beta_2 n} + \ldots + B_m(r_\perp) e^{\beta_m n} \qquad (18)$$

with the same number of positive exponents $\beta_s$. Squaring (16) gives $m^2$ terms that increase as $\exp(\gamma_i n + \gamma_j n)$, and the only possibility to have $m$ terms in (18) suggests averaging to zero for all terms with $i \neq j$. The terms with $i = j$ are positive and cannot vanish in the course of averaging:

$$\langle \psi_n^2(r_\perp) \rangle = \langle [A_1 h_n^{(1)}(r_\perp)]^2 e^{2\gamma_1 n} \rangle + \langle [A_2 h_n^{(2)}(r_\perp)]^2 e^{2\gamma_2 n} \rangle + \ldots \qquad (19)$$
$$+ \langle [A_m h_n^{(m)}(r_\perp)]^2 e^{2\gamma_m n} \rangle.$$

The terms in (18) and (19) are in one-to-one correspondence and relation between $\gamma_s$ and $\beta_s$ can be discussed for a pure 1D system when (16) and (18) have only one term in the right-hand side:

$$\psi_n \sim e^{\gamma n}, \quad \langle \psi_n \rangle \sim 1, \quad \langle \psi_n^2 \rangle \sim e^{\beta n}. \qquad (20)$$

Finiteness of $\langle \psi_n \rangle$ is insignificant in comparison with the exponential growth, and we accept $\langle \psi_n \rangle = 0$. In fact, we should discuss the usual relation between a typical value of a random quantity $x$ and its root mean square value. If $\langle x \rangle = 0$ and $\langle x^2 \rangle = \sigma^2$, then the typical value of $x$ should not be necessarily of order $\sigma$: one can state only that $|x| \lesssim \sigma$. Indeed, according to the Chebyshev inequality [49], the probability for $|x|$ to be greater than $x_0$ is less then $\sigma^2/x_0^2$. Values of $x$ substantially greater than $\sigma$ are improbable and $\sigma$ gives the upper estimate of the distribution width. The lower estimate does not exist in any form. Indeed, if distribution $P(x)$ changes on a scale of $x \sim 1$ and has a tail of $|x|^{-\alpha}$ with $1 < \alpha < 3$, then typical $x \sim 1$, while $\langle x^2 \rangle = \sigma^2 = \infty$. It is clear from these considerations that the following relation holds for the exponents in Eq. (20),

$$\beta \geq 2\gamma, \qquad (21)$$

and there are no grounds for equality.

In fact, the relation between $\beta$ and $\gamma$ can be discussed more constructively, because $\psi_n$ has a log-normal distribution [50]: i.e., the quantity $\tau = \ln|\psi_n|$ has a Gaussian distribution

$$P(\tau) \sim \exp\left\{-\frac{(\tau - an)^2}{2bn}\right\}, \qquad (22)$$

where the first and the second moments grow linearly in $n$. It is easy to see that

$$\psi_n \sim e^{an}, \quad \langle \psi_n^2 \rangle \sim e^{(2a + 2b)n}, \qquad (23)$$

and (21) obviously holds. In the 1D Anderson model we have for $E = 0$

$$a = b = \frac{1}{8}\langle V^2 \rangle, \quad \beta = \frac{1}{2}\langle V^2 \rangle \qquad (24)$$

for weak disorder, and

$$\gamma = a = \langle \ln|V| \rangle, \quad \beta = \ln\langle V^2 \rangle \qquad (25)$$

for strong disorder. For a typical distribution, Eq. (25) suggests that $b \ll a$. Analogous results are valid for many models, and situation $b \lesssim a$ should be considered typical. In this case, $\beta \sim \gamma$ and $1/\beta$ gives the correct estimate of the correlation length $\xi_{1D}$.

The situation $b \gg a$ can be discussed for an extremal case when $a = 0$, $b > 0$. Then $\psi_n$ has no systematic growth but has rare peaks with increasing amplitude (Fig. 2a). Then Mott's construction of an 1D eigenfunction gives the typical "hybrid" state, which is a linear combination of localized and extended states (Fig. 2b). The length of the localized component is evidently on the order of $1/\beta$. Consequently, an exponent $\beta$ provides essential information which is not present in the mean value of $\gamma$. This information can have a practical meaning: parameter $b$ determines the growth of all low even moments ($\langle \psi_n^{2m} \rangle \sim \exp(2ma + 2m^2b)n$), while the fourth moments enter the Kubo–Greenwood formula for conductivity.

According to numerical studies [11], an arbitrary ($s$th) term in Eq. (16) has a distribution of type (22) with parameters $a_s$ and $b_s$. Therefore, relations $\gamma_s = a_s$, $\beta_s = 2a_s + 2b_s$ hold for arbitrary $s$. We see that the second moments of $\psi_n(r_\perp)$ give valuable information: (a) exponents $\beta_s$ provide a rigorous upper bound for $\gamma_s$, $\beta_s \geq 2\gamma_s$; (b) estimates $\beta_s \sim \gamma_s$ are valid in the typical case $a_s \gtrsim b_s$; (c) $\beta_s$ are related to fluctuations of $\gamma_s$ in the case $a_s \ll b_s$. As for the heuristic relation with the Anderson transition, the use of the minimal exponents $\gamma_{\min}$ and $\beta_{\min}$ is on the same grounds. For example, scaling relations

$$\frac{1}{\gamma_{\min} L} = F\left(\frac{L}{\xi}\right) \quad \text{and} \quad \frac{1}{\beta_{\min} L} = F\left(\frac{L}{\xi}\right) \quad (26)$$

can be postulated on the same level of rigorousness. In [11], empirical scaling is stated for $\gamma_{\min} L$ and $b/a$ simultaneously. If this statement is taken seriously, it justifies (26) for $\beta_{\min} L$; in fact, scaling is absent for all these quantities (Sections 3–5).

## 3. SECOND MOMENTS FOR A SOLUTION OF THE CAUCHY PROBLEM

The idea of the present approach can be found in [51]. Let us rewrite the Schrödinger equation (11) for the 1D Anderson model as a recurrence relation, expressing $\psi_{n+1}$ in terms of $\psi_n$, $\psi_{n-1}$ and consider the Cauchy problem with the fixed initial conditions for $\psi_1$ and $\psi_0$. It is easy to see that $\psi_2$ is a function of $V_1$, $\psi_3$ is a function of $V_2$, $V_1$, etc. So $\psi_n$ and $V_n$ are statistically independent and can be averaged separately:

$$\langle \psi_{n+1} \rangle = E \langle \psi_n \rangle - \langle \psi_{n-1} \rangle, \quad (27)$$

$$\langle \psi_{n+1}^2 \rangle = (W^2 + E^2)\langle \psi_n^2 \rangle - 2E\langle \psi_n \psi_{n-1} \rangle + \langle \psi_{n-1}^2 \rangle \text{ etc.} \quad (28)$$

We have accepted that $V_n$ are statistically independent and

$$\langle V_n \rangle = 0, \quad \langle V_n V_{n'} \rangle = W^2 \delta_{nn'}. \quad (29)$$

Equation (27) has the form (11) with $V_n \equiv 0$ and its solutions are

$$\langle \psi_n \rangle \sim \exp(ipn) \quad (30)$$

with $2\cos p = E$. Inside the allowed band, they have no systematic growth and $\langle \psi_n \rangle \sim 1$. Equation (28) for $E = 0$ is a difference equation for $x_n = \langle \psi_n^2 \rangle$,

$$x_{n+1} = W^2 x_n + x_{n-1} \quad (31)$$

with exponential solution

$$x_n = \langle \psi_n^2 \rangle \sim e^{\beta n}, \quad 2\sinh\beta = W^2. \quad (32)$$

In the case $E \neq 0$, Eq. (28) is not closed and should be completed by the equation

$$\langle \psi_{n+1} \psi_n \rangle = E \langle \psi_n^2 \rangle - \langle \psi_n \psi_{n-1} \rangle. \quad (33)$$

As a result, a set of difference equations arises for $x_n = \langle \psi_n^2 \rangle$ and $\psi_n = \langle \psi_n \psi_{n-1} \rangle$

$$\begin{aligned} x_{n+1} &= (W^2 + E^2)x_n + x_{n-1} - 2Ey_n, \\ y_{n+1} &= Ex_n - y_n, \end{aligned} \quad (34)$$

with exponential growth of solution.

This approach is easily generalized for an arbitrary quasi-1D system. Consider the 2D Anderson model

$$\psi_{n+1,m} + \psi_{n-1,m} + \psi_{n,m+1} + \psi_{n,m-1} + V_{n,m}\psi_{n,m} = E\psi_{n,m} \quad (35)$$

and interpret it as a recurrence relation in $n$. Solving (35) for the quantity $\psi_{n+1,m}$ and averaging its square, we can express it via the pair correlators of $\psi_{n,m}$ containing lower values of $n$. Constructing analogous equations for other correlators, we end with the close system of difference equations for the quantities

$$\begin{aligned} x_{m,m'}(n) &\equiv \langle \psi_{n,m}\psi_{n,m'} \rangle, \\ y_{m,m'}(n) &\equiv \langle \psi_{n,m}\psi_{n-1,m'} \rangle, \\ z_{m,m'}(n) &\equiv \langle \psi_{n-1,m}\psi_{n,m'} \rangle, \end{aligned} \quad (36)$$

which for $E = 0$ has the form

$$\begin{aligned} x_{m,m'}(n+1) &= W^2 \delta_{m,m'} x_{m,m'}(n) + x_{m+1,m'+1}(n) \\ &+ x_{m-1,m'+1}(n) + x_{m+1,m'-1}(n) + x_{m-1,m'-1}(n) \\ &+ x_{m,m'}(n-1) + y_{m+1,m'}(n) + y_{m-1,m'}(n) \\ &+ z_{m,m'+1}(n) + z_{m,m'-1}(n), \end{aligned} \quad (37)$$

$$y_{m,m'}(n+1) = -x_{m+1,m'}(n) - x_{m-1,m'}(n) - z_{m,m'}(n),$$

$$z_{m,m'}(n+1) = -x_{m,m'+1}(n) - x_{m,m'-1}(n) - y_{m,m'}(n).$$

This is a set of linear equations with coefficients independent of $n$, and its solution is exponential in $n$ [52]:

$$x_{m,m'}(n) = x_{m,m'}e^{\beta n}, \quad y_{m,m'}(n) = y_{m,m'}e^{\beta n}, \quad z_{m,m'}(n) = z_{m,m'}e^{\beta n}. \tag{38}$$

The formal change of variable is useful

$$x_{m,m'} \equiv \tilde{x}_{m,m'-1} \equiv \tilde{x}_{m,l} \quad \text{etc.}, \tag{39}$$

where $l = m' - m$. Then, we have (with tildes omitted)

$$(e^\beta - e^{-\beta})x_{m,l} = W^2 \delta_{l,0} x_{m,l} + x_{m+1,l}$$
$$+ x_{m-1,l} + x_{m+1,l-2} + x_{m-1,l+2}$$
$$+ y_{m+1,l-1} + y_{m-1,l+1} + z_{m,l+1} + z_{m,l-1}, \tag{40}$$

$$e^\beta y_{m,l} = -x_{m+1,l-1} - x_{m-1,l+1} - z_{m,l},$$

$$e^\beta z_{m,l} = -x_{m,l+1} - x_{m,l-1} - y_{m,l}.$$

The coefficients contain no $m$ dependence, and solution is exponential in $m$:

$$x_{m,l} = x_l e^{ipm} \quad \text{etc.}, \tag{41}$$

where allowed values for $p$, $p_s = 2\pi s/L$, $s = 0, 1, \ldots, L - 1$ are determined by the periodical boundary conditions in the transverse direction:

$$\psi_{n,m+L} = \psi_{n,m}. \tag{42}$$

Excluding $y_{m,l}$ and $z_{m,l}$ from the first equation in (40), we end with the equation

$$x_{l+2}e^{-ip} + x_{l-2}e^{ip} + V\delta_{l,0}x_l = \epsilon x_l, \quad x_{l+L} = x_l,$$
$$\epsilon = 2\cosh\beta, \quad V = \frac{W^2 \sinh\beta}{\cosh\beta - \cos p}, \tag{43}$$

describing a single impurity in a periodic chain. For $L \to \infty$ its solution has the form $x_l \sim \exp(ipl/2 - \beta|l|/2)$ and the initial correlator

$$\langle \psi_{n,m} \psi_{n,m'} \rangle$$
$$\sim \exp\left\{ ip\frac{m+m'}{2} - \beta\frac{|m-m'|}{2} + \beta n \right\} \tag{44}$$

is localized in the transverse direction on the same scale $1/\beta$, as the scale of its growth in $n$. As a result, the localization length for the 2D system coincides with $\xi_{1D}$.

The positive exponents $\beta_s$ for finite odd $L$ are determined by the equation

$$2(\cosh\beta_s - \cos p_s) = W^2 \coth(\beta_s L/2), \quad p_s = 2\pi s/L, \quad s = 0, 1, \ldots, L-1. \tag{45}$$

Their number is equal to $L$ and coincides with a number of positive Lyapunov exponents $\gamma_s$ for the same problem.[5] Allowed values of $p_s$ and $\beta_s$ become dense in the large $L$ limit, and the quantities $\beta$ and $p$ can be considered as continuous:

$$2(\cosh\beta - \cos p) = W^2 \coth(\beta L/2). \tag{46}$$

The minimal value of $\beta$ is realized for $p = \pi$ and can be easily found in the large $L$ limit:

$$\beta_{\min} = \begin{cases} \operatorname{arccosh}(W^2/2 - 1), & W^2 > 4 \\ \dfrac{2}{L}\operatorname{arctanh}(W^2/4), & W^2 < 4 \\ \dfrac{2\ln L - 2\ln\ln L + \ldots}{L}, & W^2 = 4. \end{cases} \tag{47}$$

The character of the solution is qualitatively changed at the critical value $W_c = 2$. If $W > W_c$, Eq. (46) is solved for $\beta \sim 1$, $\beta L \to \infty$ and $\beta_{\min}$ tends to a constant in the large $L$ limit. If $W < W_c$, Eq. (46) has solution for $\beta L = \text{const}$, $\beta \to 0$ and provides for the behavior[6] $\beta_{\min} \propto 1/L$ for $L \to \infty$. If $W = W_c$, solution is sought at conditions $\beta L \gg 1$, $\beta \ll 1$, when Eq. (46) reduces to $\beta^2 = 8\exp(-\beta L)$ and can be solved iteratively.

If the correlation length $\xi_{1D}$ is estimated as $1/\beta_{\min}$, comparison with Section 2 leads to the conclusion that a state with the long-range order (i.e., the metallic phase) is absent. Exponential localization takes place for $W > W_c$, while the critical behavior $\xi_{1D} \sim L$ is realized in the entire range of $W < W_c$. The latter situation corresponds to localization with the divergent correlation length $\xi_{\text{loc}} \sim L$ and should probably be interpreted as power law localization. The transition at $W = W_c$ is of the Kosterlitz–Thouless type and should not be confused with the usual Anderson transition.

---

[5] The matrix $T$ in Eq. (15) has dimensions $2L \times 2L$, but its eigenvalues occur in pairs of $e^{\gamma_s}$ and $e^{-\gamma_s}$, so the number of positive $\gamma_s$ is equal to $L$. In the case of even $L$, the number of positive $\beta_s$ does not coincide with $L$ and there are difficulties in comparing (16) and (18).

[6] Vanishing of $\beta_{\min}$ for $L = \infty$ was obtained in [42].

Calculating the first corrections to (47) related with finiteness of $L$, we have for $W > W_c$

$$\beta_{min} = \beta_\infty + \frac{W^2}{\sinh\beta_\infty} e^{\beta_\infty L},$$
$$\beta_\infty = \text{arccosh}\left(\frac{W^2 - 2}{2}\right)$$
(48)

and for $W < W_c$,

$$g = \frac{1}{\beta_{min} L} = g_\infty + \frac{2(\sinh 1/2 g_\infty)^2}{W^2 L^2},$$
$$g_\infty = \frac{1}{2\,\text{arctanh}(W^2/4)}.$$
(49)

Defining the correlation length $\xi$ as a scale, where dependences (48), (49) reach their asymptotics (i.e., where the additional terms become comparable to the main terms), we have (Fig. 3)

$$\xi \sim \begin{cases} 1/\ln W^2, & W^2 \longrightarrow \infty \\ \dfrac{\ln(1/\tau)}{\sqrt{\tau}}, & \tau = |W - W_c| \longrightarrow 0 \\ W^2, & W^2 \longrightarrow 0. \end{cases}$$
(50)

If $W > W_c$, the correlational length $\xi$ coincides with the localization length $\xi_{loc} \sim \xi_{1D}$ apart from the logarithmic corrections. If $W < W_c$, the scales $\xi$ and $\xi_{loc}$ are substantially different, as is typical for the metallic phase (Fig. 3).

The scaling parameter $g(L)$ can be defined as $1/\beta_{min} L$. Its dependence on $L$ is determined by the equation

$$2\cosh\frac{1}{gL} - 2\cos p = W^2 \cot\frac{1}{2g}$$
(51)

with $p = \pi$ and presented in Fig. 1b. One can see the significant difference from the typical scaling situation (Fig. 1a). Absence of one-parameter scaling in Fig. 1b is clear from the fact that $g(L)$ is not constant for $W = W_c$, as it should be according (8), (9). It is still more evident for $W < W_c$, when different curves have different constant limits for $L \longrightarrow \infty$ and certainly cannot be matched by a scale transformation.

In the above considerations, we have estimated $\xi_{1D}$ as $1/\beta_{min}$. This can arouse doubts, because in the absence of scaling the quantities $\beta_{min}$ and $\gamma_{min}$ can be very different. In fact, substitution $\beta_{min}$ by $\gamma_{min}$ does not lead to qualitative changes in the presented picture. Indeed, $\beta_{min}$ provides a rigorous upper bound for $\gamma_{min}$ and (47) leads to

$$\gamma_{min} \longrightarrow 0 \text{ for } L \longrightarrow \infty, \text{ if } W < W_c.$$
(52)

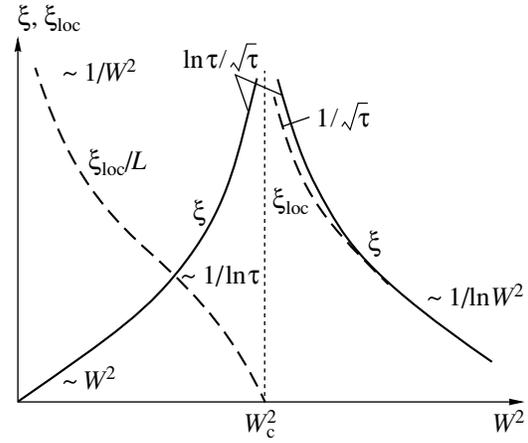

**Fig. 3.** Characteristic scales $\xi$ and $\xi_{loc}$ obtained under the assumption $\xi_{1D} \sim 1/\beta_{min}$.

This is sufficient for the existence of the 2D transition, because in the large $W$ region, the existence of exponential localization is beyond any doubt and finiteness of $\gamma_{min}$ has been reliably established by numerical investigations [8–11]. Of course, the upper bound for $\gamma_{min}$ does not forbid it to decrease more rapidly than $1/L$, as it should be for a true metallic state. However, such a possibility is reliably excluded by numerical studies (Section 4). Nevertheless, substitution $\gamma_{min}$ by $\beta_{min}$ can change the position of the critical point and the character of the critical behavior. Thus, the presented quantitative results should be considered as illustrative.

The influence of the phase transition on conductivity can be seen from the following arguments. Conductance $G$ of a quasi-1D system of length $l$ is roughly given by the exponential $\exp\{-2\gamma_{min} l\}$ (see [11] and references therein). Extrapolation to $l \sim L$ suggests that $G \sim \exp\{-\text{const} L\}$ for $W > W_c$, while for $W < W_c$ the exponential reduces to a constant (in view of $\gamma_{min} \sim 1/L$) and dependence $G(L)$ is determined by a preexponential factor.

## 4. COMPARISON WITH NUMERICAL RESULTS

The idea of power law localization was put forward by Last and Thouless [53] and discussed in a number of papers [54]. The statements, literally coinciding with those of Section 3, were made by Pichard and Sarma in 1981 [7] as the result of a numerical study of the 2D Anderson model. Their dependences of $\xi_{1D}$ on $L$ are presented in Fig. 4a, where values of disorder correspond to the quantity

$$\tilde{W} = W\sqrt{12}$$
(53)

(so $\tilde{W}_c = \sqrt{48} = 6.928...$), because a rectangular distribution of width $\tilde{W}$ was used for $V_n$ with $\langle V^2 \rangle =$ $\tilde{W}^2/12 \equiv W^2$. The dependences are clearly linear for $\tilde{W} < 6$, while a tendency to saturation arises for $\tilde{W} > 6$ with a clear saturation for large $\tilde{W}$.

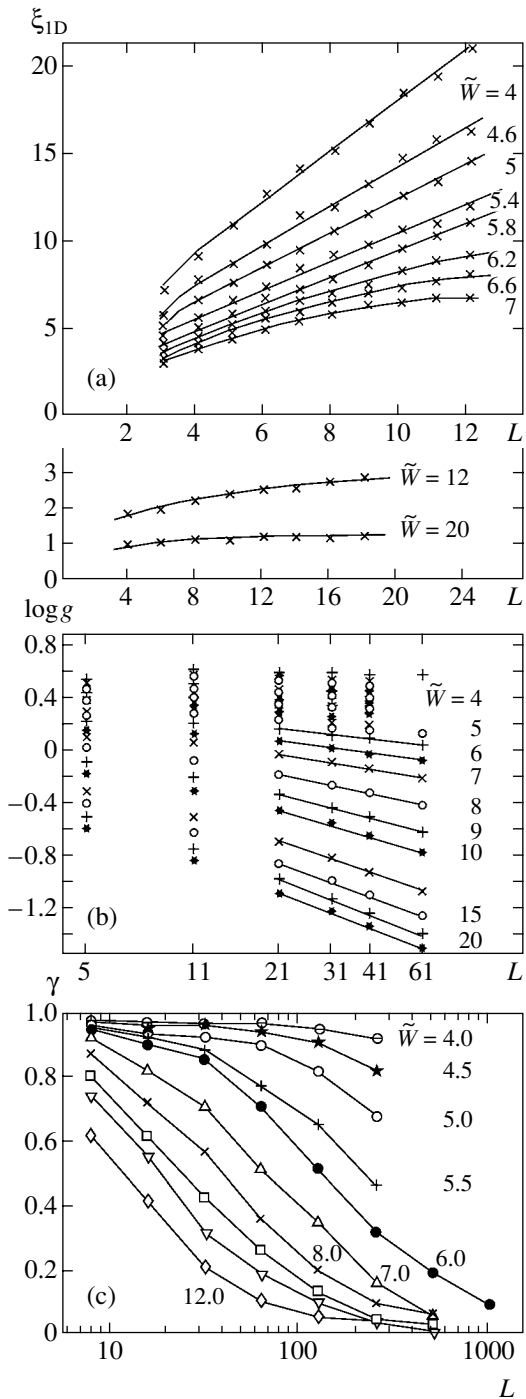

**Fig. 4.** Numerical results for $\xi_{1D}$ [7], $g = 1/\gamma_{min}L$ [10] and parameter $\gamma(L)$ related to the energy level statistics [13].

The results of [7] are considered out of date [8–11], and it is instructive to analyze the raw data of [10], which are cited as the best in the context of the transfer matrix method (Fig. 4b). One can see that the scaling parameter $g = 1/\gamma_{min}L$ decrease convincingly only for large $\tilde{W}$. In the range of intermediate disorder ($\tilde{W} = 4$–7), one cannot say definitely whether is there a tendency to unbounded decrease or to saturation. The data for weak disorder ($\tilde{W} < 4$) are absent altogether.

Thus, the raw numerical data (Fig. 4b) do not demonstrate absence of the 2D phase transition as stated by the authors of [10]. The latter conclusion is based on interpretation of these data in terms of one-parameter scaling. However, such interpretation is surely invalid. Absence of scaling for $\beta_{min}$ suggests absence of scaling for $\gamma_{min}$, and this is confirmed by the similarity of Figs. 1b and 4b. The use of $\beta_{min}$ as the upper bound for $\gamma_{min}$ leads to the conclusion that the curves for $W < W_c$ in Fig. 4b cannot decrease to zero and should tend to finite limits. The scaling ansatz (8) can be formally valid only in the case if these finite limits are the same for all curves with $W < W_c$. Such a possibility does not appear realistic in Fig. 4b and, in fact, can be excluded: for small disorder, the lower bound given by $1/\beta_{min}L$ lies higher than all the data of Fig. 4b.

It is admitted in [8–11] that scaling relation (26) for $\gamma_{min}$ is not proved, but it is stated that this relation has been convincingly confirmed empirically. Scaling curves $g = F(L/\xi)$ of impressive quality are presented in [10]. However, one should be very careful with empirical proofs of scaling. It is possible to come up with an algorithm that allows "proof" of empirical scaling in practically any situation.

Let us discuss construction of scaling curves in more detail. The raw numerical data are represented by dependences $g(L)$ for fixed values of disorder $W_0$, $W_1$, $W_2$,… (Fig. 5a). They should be plotted in Fig. 5b as functions of $L/\xi$, where the value of $\xi$ for each curve should be chosen in such a manner that all curves coincide. If the logarithmic scale is chosen along the $L$ axis, this procedure reduces to simple translation. Let the curve for $W_0$ be taken as a reference and a corresponding value $\xi_0$ be accepted as unity. Then this curve is carried over to Fig. 5b without changes. Now the curve for $W_1$ is translated to match the curve for $W_0$, a corre-sponding value $\xi_1$ is determined, and so on. In the logarithmic plot, the dependences $g(L)$ have a simple form (see Fig. 4b) and can be approximated by something like parabolas. If the values $W_0$ and $W_1$ are close, the corresponding curves are parabolas with slightly different coefficients and they fit sufficiently well after translation.

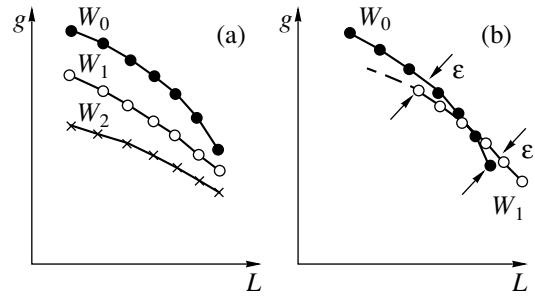

**Fig. 5.** Construction of scaling curves in a situation with no scaling.

Let us take $\epsilon$ as the permissible error of fitting and superimpose the curve for $W_1$ in a crosswise manner (Fig. 5b) on the curve for $W_0$. If some part of the former curve does not fit sufficiently well (dotted line in Fig. 5b), the corresponding points can be debated on reasonable grounds: scaling is a large-scale property and the raw data for small $L$ are not reliable. The curve for $W_2$ is superimposed analogously, etc. If there is sufficient scattering of points, such a procedure will look natural. If the scattering of points is small, one can take a small step increment in $W$: then numerous curves will densely fill a band of width $\epsilon$ and the resulting scaling curve will appear accurate.

One can see that it is rather difficult to recognize a situation with no scaling from a situation when scaling holds but there are significant corrections to it. In the case under consideration, the situation is close to scaling in the sense that scaling relation (26) is trivially valid for $L \gtrsim \xi$ in the localized phase, when $\beta_{\min} \approx 1/\xi$ and $g \approx \xi/L$ in correspondence with Eq. (26) for $F(x) \approx 1/x$.

Certain comments should be made on the variant of finite-size scaling based on the level statistics [12]. In this case, rather large systems are used, up to $1024^2$ [13], and localization of all states in 2D systems appear convincing on the level of raw data (Fig. 4c), without interpreting them in terms of one-parameter scaling. However, this approach deals with crossover between the metallic behavior at small $L$ and localized behavior at large $L$, and no attempt has been made to distinguish between exponential and power law localization.

## 5. IS ONE-PARAMETER SCALING POSSIBLE?

In Section 3 we have shown violation of one-parameter scaling for the quantity $\beta_{\min}$. If $\beta_{\min} \sim \gamma_{\min}$, then scaling is absent also for $\gamma_{\min}$. If $\beta_{\min}$ and $\gamma_{\min}$ are essentially different, a quasi-1D eigenfunction has a structure corresponding to both these parameters (see Fig. 2) and scaling is impossible on physical grounds. Analysis of numerical data (Section 4) confirms these conclusions. Two possible conclusions can be derived:

(i) the one-parameter scaling hypothesis [1] is fundamentally wrong;

(ii) the minimal Lyapunov exponent is an incorrect scaling variable.

Possibility (i) is not as absurd as it seems. Justifications for scaling in the $\sigma$-model approach [33] in fact failed due to a high-gradient catastrophe [55, 56], and absence of scaling on the level of distribution func-tions [57] set a problem of the correct choice of scaling variables. As for experiment, it confirms the "theory of quantum corrections" rather than the "theory of weak localization." Nevertheless, we consider the one-parameter scaling hypothesis as physically convincing. Its validity is confirmed (a) by analytical scaling in quasi-random systems [58–60]; (b) by validity of scaling relation $s = \nu(d-2)$ in the Vollhardt and Wölfle type theories [25, 26, 41]; (c) by renormalizability for $d < 4$ in the exact field theoretical formulation of the problem [29–31] (see footnote 1).

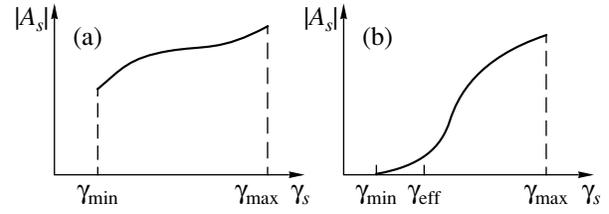

**Fig. 6.** Coefficients $A_s$ in Eq. (13), appearing in Mott's construction, as function of $\gamma_s$.

Let us consider possibility (ii). It is clear from Section 2 that the existence of scaling for the correlation length $\xi_{1D}$ looks convincing, and this is confirmed by the experience of the phase transitions theory [44–46]. As for relation (10) between $\xi_{1D}$ and $\gamma_{\min}$, this is not as evident as it seems: for example, situations with $\gamma_{\min} > 0$ and $\xi_{1D} = \infty$ are known in quasi-random systems [58, 59, 61].[7] Relation (10) is based on Mott's construction of eigenfunctions by matching two solutions of type (16), which increase from two opposite edges of the system. Exact matching needs all terms in Eq. (16), and consequently, the coefficient $A_m$ is finite, providing a length scale related to $1/\gamma_{\min}$. These considerations are valid for a sufficiently small number of terms in Eq. (16). In the large $L$ limit, a spectrum of the Lyapunov exponents becomes quasi-continuous and a number of terms in Eq. (16) tend to infinity. In such a situation, no particular term in Eq. (16) is essential: it is an integral effect from all terms corresponding to some interval of the spectrum $\gamma_s$ that matters.

Let us consider the coefficients $A_s$ in Eq. (16), appearing in Mott's construction, as a function of $\gamma_s$ (Fig. 6). Two qualitatively different situations are possible. In the first of them (Fig. 6a), all $A_s$ are of the same order of magnitude; then the vicinity of $\gamma_{\min}$ makes a significant contribution and the length scale $1/\gamma_{\min}$ indeed corresponds to the localization length $\xi_{1D}$. In the other situation (Fig. 6b), the contribution of the vicinity of $\gamma_{\min}$ is strongly suppressed and the length scale $1/\gamma_{\min}$ has no physical meaning. The localization length $\xi_{1D}$ is determined by some effective exponent $\gamma_{\text{eff}}$ that provides a suitable cutoff[8] in the small $\gamma_s$ region (Fig. 6b)

$$\xi_{1D} \sim \frac{1}{\gamma_{\text{eff}}}. \qquad (54)$$

---

[7] In the 1D model (11) with $V_n = V\cos(2\pi\beta n)$ and irrational $\beta$, the Anderson transition holds for $V = 2$ [58, 59, 61]. The Lyapunov exponent $\gamma$ is positive for all irrational $\beta$ in the $V > 2$ region [61]. Nevertheless, localization length diverges for certain values of $\beta$, which are anomalously close to rational numbers [59].

According to (8), scaling relation (26) should be postulated not for $\gamma_{\min}$ but for $\gamma_{\text{eff}}$.[9] After that, the dependence of $\gamma_{\text{eff}}$ on parameters is determined by scaling itself.

The latter statement can be easily demonstrated in the framework of the numerical algorithm. Let us return to Fig. 5 but now accept that the curves for $W_0$, $W_1$, ... are related not to $g = 1/\gamma_{\min} L$ but to $g = 1/\gamma_{n_0} L$ where $\gamma_{n_0}$ is the Lyapunov exponent with the fixed number $n_0$. In general, the curves for $W_0$ and $W_1$ cannot be fit well by a scale transformation. One can improve the situation in the following manner. Taking a step from $W_0$ to $W_1$, let us replace $\gamma_{n_0}$ by $\gamma_{n_1}$, i.e., change the number of the Lyapunov exponent. The curve for $W_1$ will change its form and we can choose $\Delta n = n_1 - n_0$ from the condition of the best fit with the curve for $W_0$. Analogously, for $W_2$ we take $\gamma_{n_2}$, adjust $n_2$, and so on. As a result, the scaling construction will determine not only dependence $\xi(W)$ but also dependence $n(W)$. Of course, these dependences are not determined completely: the general scale for $\xi$ and initial number $n_0$ remain arbitrary.

Thus we came to the constructive modification of the commonly accepted numerical algorithm. This modification makes it possible to improve the quality of scaling and will probably resolve the contradictions discussed in Section 1.

## 6. ANALYTICAL SCALING

The suggested algorithm can be realized analytically if the raw data are given in the form defining $g = 1/\gamma_p L$ as a function of $p$, $W$, $L$:

$$g = Q(p, W, \tau), \quad \tau = \ln L, \qquad (55)$$

where $p$ is a continuous number of the Lyapunov exponent analogous to that in Eq. (51). Linearization of (55) near some value $\tau_0$ gives

$$\begin{aligned} g &= Q(p, Q, \tau_0) + Q'_\tau(p, W, \tau_0)(\tau - \tau_0) \\ &\equiv g_0 + A(\tau - \tau_0). \end{aligned} \qquad (56)$$

Analogous linearization for $W_1$ near the value $\tau_1$ chosen from the condition $Q(p, W_1, \tau_1) = Q(p, W, \tau_0)$

$$\begin{aligned} g &= Q(p, W_1, \tau_1) + Q'_\tau(p, W_1, \tau_1)(\tau - \tau_1) \\ &\equiv g_0 + B(\tau - \tau_1) \end{aligned} \qquad (57)$$

gives a slope $B$ different from $A$, and the linear portions of dependences (56), (57) cannot be matched by a scale transformation. Let us change $p$ in Eq. (57) in such way that equality $A = B$ holds:

$$\begin{aligned} g &= Q(p_1, W_1, \tau_1) + Q'_\tau(p_1, W_1, \tau_1)(\tau - \tau_1) \\ &\equiv g_0 + A(\tau - \tau_1). \end{aligned} \qquad (58)$$

If $p_1$, $W_1$, $\tau_1$ are close to $p$, $W$, $\tau_0$, then correspondence of (56) to (58) gives

$$\begin{aligned} Q'_p(p, W, \tau_0)\Delta p &+ Q'_W(p, W, \tau_0)\Delta W \\ &+ Q'_\tau(p, W, \tau_0)\Delta\tau = 0, \\ Q''_{\tau p}(p, W, \tau_0)\Delta p &+ Q''_{\tau W}(p, W, \tau_0)\Delta W \\ &+ Q''_{\tau\tau}(p, W, \tau_0)\Delta\tau = 0, \end{aligned} \qquad (59)$$

or solving for $\Delta p$ and $\Delta\tau$,

$$\begin{aligned} \Delta p &= \frac{Q'_\tau Q''_{\tau W} - Q'_W Q''_{\tau\tau}}{Q'_p Q''_{\tau\tau} - Q'_\tau Q''_{\tau p}} \Delta W, \\ \Delta\tau &= \frac{Q'_W Q''_{\tau p} - Q'_p Q''_{\tau W}}{Q'_p Q''_{\tau\tau} - Q'_\tau Q''_{\tau p}} \Delta W. \end{aligned} \qquad (60)$$

If an increment of $\tau$ is interpreted as an increment of $\ln\xi$,

$$\Delta\tau = \Delta\ln\xi, \qquad (61)$$

then Eq. (58) takes the form

$$\Delta g = Q'_\tau(p, W, \tau_0)\Delta\ln(L/\xi). \qquad (62)$$

For infinitesimal increments, Eqs. (60)–(62) turn into a set of the differential equations

---

[8] It is evident from Eq. (44) that the $s$th term of Eq. (18) is localized in the transverse direction on the same scale $1/\beta_s$ as the scale of its growth in $n$. An analogous property is expected for Eq. (16) and provides the equality of the transverse and longitudinal correlation lengths. The latter fundamental property is not spoiled when the lower edge of spectrum $\gamma_{\min}$ is replaced by the effective cutoff $\gamma_{\text{eff}}$.

[9] A quasi-1D eigenfunction contains a lot of scales $1/\gamma_1$, $1/\gamma_2$, ..., $1/\gamma_m$ and all these scales are essential near its center. Small scales succeedingly "die out" when one moves from the maximum of the eigenfunction to its tails. Only scale $1/\gamma_m$ remains in the end, but for the situation of Fig. 6b it occurs at such distances where the eigenfunction is zero for all practical purposes. A single parameter $\xi_{1D}$ cannot adequately describe all scales $1/\gamma_s$. In the best case, it can account for the most significant of them, those which determine the general form of the eigenfunction and correspond to the effective cutoff $\gamma_{\text{eff}}$.

$$\frac{dp}{dW} = -\frac{Q'_W(p,W,\tau_0)Q''_{\tau\tau}(p,W,\tau_0) - Q'_\tau(p,W,\tau_0)Q''_{\tau W}(p,W,\tau_0)}{Q'_p(p,W,\tau_0)Q''_{\tau\tau}(p,W,\tau_0) - Q'_\tau(p,W,\tau_0)Q''_{\tau p}(p,W,\tau_0)},$$

$$\frac{d\ln\xi}{dW} = -\frac{Q'_p(p,W,\tau_0)Q''_{\tau W}(p,W,\tau_0) - Q'_W(p,W,\tau_0)Q''_{\tau p}(p,W,\tau_0)}{Q'_p(p,W,\tau_0)Q''_{\tau\tau}(p,W,\tau_0) - Q'_\tau(p,W,\tau_0)Q''_{\tau p}(p,W,\tau_0)}, \quad (63)$$

$$\frac{dg}{d\ln(L/\xi)} = Q'_\tau(p,W,\tau_0),$$

defining the dependences $p(W)$, $\xi(W)$ and $g = F(L/\xi)$. Equations (63) correspond to the usual scaling construction (Sections 4 and 5) for the maximal system size $L_0 = \exp(\tau_0)$, where dependences $g(\ln L)$ are linearized near $\ln L_0$ and only linear portions (marked in Fig. 4b) are matched in the course of scale transformations. The $\tau_0$ dependence should vanish in the limit $\tau_0 \to \infty$ for the approach to be self-consistent.

If the dependence (55) is given in the implicit form

$$G(g, p, W, \tau) = 0, \quad (64)$$

then Eqs. (63) can be expressed in terms of $G$:

$$\frac{dp}{dW} = -\frac{G'_W(G''_{g\tau}G'_\tau - G''_{\tau\tau}G'_g) - G'_\tau(G''_{gW}G'_\tau - G''_{\tau W}G'_g)}{G'_p(G''_{g\tau}G'_\tau - G''_{\tau\tau}G'_g) - G'_\tau(G''_{gp}G'_\tau - G''_{\tau p}G'_g)},$$

$$\frac{d\ln\xi}{dW} = -\frac{G'_p(G''_{gW}G'_\tau - G''_{\tau W}G'_g) - G'_W(G''_{gp}G'_\tau - G''_{\tau p}G'_g)}{G'_p(G''_{g\tau}G'_\tau - G''_{\tau\tau}G'_g) - G'_\tau(G''_{gp}G'_\tau - G''_{\tau p}G'_g)}, \quad (65)$$

$$\frac{dg}{d\ln(L/\xi)} = -\frac{G'_\tau}{G'_g}.$$

All quantities in the right-hand side are functions of $g_0, p, W, \tau_0$, where $\tau_0$ is a constant parameter and $g_0$ is expressed in terms of $p, W$ using the relation $G(g_0, p, W, \tau_0) = 0$.

Unfortunately, relation (64) for the conventional Lyapunov exponents $\gamma_s$ is not available; therefore, we present here illustrative calculations for the exponents $\beta_s$ when Eq. (64) has the form (51). The latter equation can be simplified by expansion of $\cosh(1/gL)$ without significant physical consequences.[10] Thus, Eq. (64) can be taken in the form

$$G(g, p, W, \tau) = \frac{\exp(-2\tau_0)}{g^2} + \varphi(p) - W^2 f(g) = 0, \quad (66)$$

where

$$\varphi(p) = 2(1 - \cos p), \quad f(g) = \coth(1/2g).$$

Then Eqs. (65) reduce to

$$\frac{d\varphi(p)}{dW^2} = -\frac{g_0^2 f(g_0)W^2 - \exp(-2\tau_0)}{g_0^2 W^2} = \frac{\varphi(p)}{W^2}, \quad (67)$$

$$\frac{d\ln\xi}{d\ln W} = -1, \quad (68)$$

$$\frac{dg}{d\ln\xi} = \frac{2g_0 \exp(-2\tau_0)}{2\exp(-2\tau_0) + W^2 g_0^3 f'(g_0)}, \quad (69)$$

where $g_0$ is a function of $p, W$ determined by equation $G(g_0, p, W, \tau_0) = 0$. It is easy to solve (67), (68)

$$\varphi(p) = 2(1 - \cos p) = c_0 W^2, \quad \xi = \frac{c_1}{W} \quad (70)$$

and obtain the relations

$$W^2 = \frac{\exp(-2\tau_0)}{g_0^2[f(g_0) - c_0]}, \quad \frac{dg}{dg_0} = 1, \quad (71)$$

which make it impossible to find the dependence $g = F(L/\xi)$ in the implicit form

$$c_1^2 g^2[f(g) - c_0] = (\xi/L)^2. \quad (72)$$

Here $c_0$ and $c_1$ are arbitrary constants. The quantity $\varphi(p)$ is restricted, $0 \le \varphi(p) \le 4$ and scaling is possible only for

$$W^2 < \frac{4}{c_0} \equiv W_{c1}^2. \quad (73)$$

---

[10] It gives only restriction for $L$ from below in the small $g$ region.

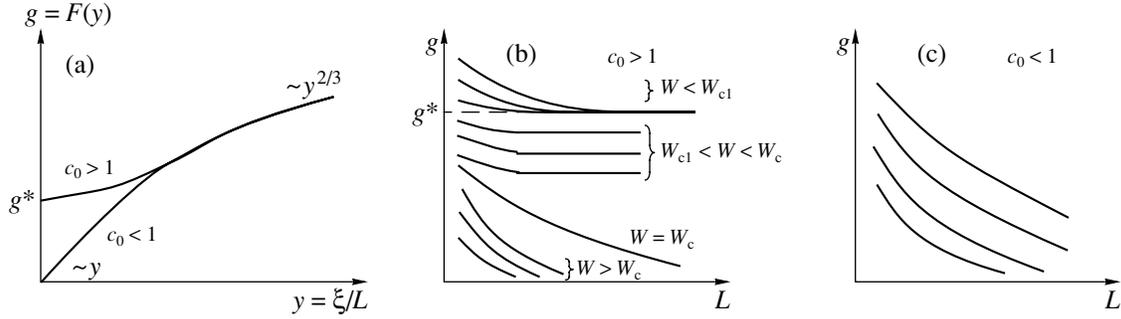

**Fig. 7.** (a) Scaling function $g = F(\xi/L)$ given by Eq. (72), (b) and (c) dependences $g(L)$ for $c_0 > 1$ and $c_0 < 1$.

If $W > W_{c1}$, the quantity $\varphi(p)$ takes its maximum value at $p = \pi$ and the concept of the minimal Lyapunov exponent is restored; therefore, scaling is absent and the results of Section 3 hold.

The situation is qualitatively different for $c_0 < 1$ and $c_0 > 1$, when $W_{c1} > W_c$ and $W_{c1} < W_c$, respectively. A scaling function $g = F(\xi/L)$ given by Eq. (77) is shown in Fig. 7a: for $c_0 < 1$ it is close to the one empirically found in [10], while for $c_0 > 1$, a finite limit $g^* = 1/2 \operatorname{arctanh}(1/c_0)$ is reached for $\xi/L \longrightarrow 0$. The dependences $g(L)$ for $W < W_{c1}$ can be found from Eq. (66) after substitution of $\varphi(p)$ from Eq. (70), while for $W > W_{c1}$ they remain the same as in Section 3 (Figs. 7b and 7c). The behavior of the characteristic scales for $c_0 > 1$ and $c_0 < 1$ is shown in Figs. 8a and 8b, respectively.

It is clear from Figs. 7 and 8 that exponential localization of all states takes place for $c_0 < 1$ in correspondence with the commonly accepted viewpoint, while the phase with power law localization remains for $c_0 > 1$, though the behavior of characteristic scales changes in comparison with Section 3. In fact, singularity at the point $W_{c1}$ is false. It is related to our postulation of exact scaling for $W < W_{c1}$, which is in fact approximate. The correlation length $\xi$ is finite near $W_{c1}$, and corrections to scaling related to $l_i/\xi$ (see Eq. (4)) cannot be considered vanishingly small. With corrections to scaling taken into account, the qualitative difference between regions $W < W_{c1}$ and $W > W_{c1}$ disappears. There is good scaling for $W \lesssim W_{c1}$ and absence of scaling for $W \gtrsim W_{c1}$, but destruction of scaling occurs gradually due to the increase of corrections to it.

Let us discuss the physical sense of an arbitrary parameter $c_0$. Formally, it occurs due to the absence of initial conditions to Eqs. (63), while in the specific Anderson model the value of $c_0$ is definite. However, we have not fixed the distribution function $P(V)$ and used only its first and second moments (see Eq. (27)). Therefore, the initial equation (66) describes not one, but a variety of Anderson models with different forms of $P(V)$. The values of $c_0$ are different in these models, and we can expect them to cover both $c_0 < 1$ and $c_0 > 1$ regions.[11] As a result, 2D systems can be divided into two classes. The first class is characterized by exponential localization of all states, while in the second class there is a phase transition between exponential and power law localization. Division into two classes was proposed by Zavaritskaya in the middle of 1980s on experimental grounds (see [62] and references therein).

We should note that the above consideration has an illustrative character. The initial Eq. (66) has another form for the conventional Lyapunov exponents, and substantial modification of the quantitative results is possible. In particularly, instead of (70), one expects the exponential dependence $\xi \sim \exp(\mathrm{const}/W^2)$ for the correlation length, as follows from one-parameter scaling [1] or from the Vollhardt and Wölfle theory [25].

## 7. INTERPRETATION IN TERMS OF THE GELL-MANN–LOW EQUATION

In one-parameter scaling theory [1], a scaling variable $g(L)$ is defined as a conductance $G_L$ of a finite block of size $L^d$ in units of $e^2/h$. The Gell-Mann–Low equation is valid for it:

$$\frac{d \ln g}{d \ln L} = \beta(g), \qquad (74)$$

where $\beta(g)$ has asymptotical behavior,

$$\beta(g) = \begin{cases} (d-2) + \dfrac{A}{g} + \ldots \quad (A < 0), \quad g \gg 1 \\ \ln g, \quad g \ll 1. \end{cases} \qquad (75)$$

The zero term of the first asymptotics is related to the existence of finite conductivity $\sigma$ in the metallic state (so $G_L \sim \sigma L^{d-2}$) and the additional term $A/g$ is obtained by a diagrammatical analysis [63]. The second asymp-totics is related to assumption of exponential localization ($G_L \sim \exp(-\text{const}\,L)$).

---

[11]Of course, there may be principal restrictions that make realization of the case $c_0 > 1$ impossible. At present, we know nothing of such restrictions.

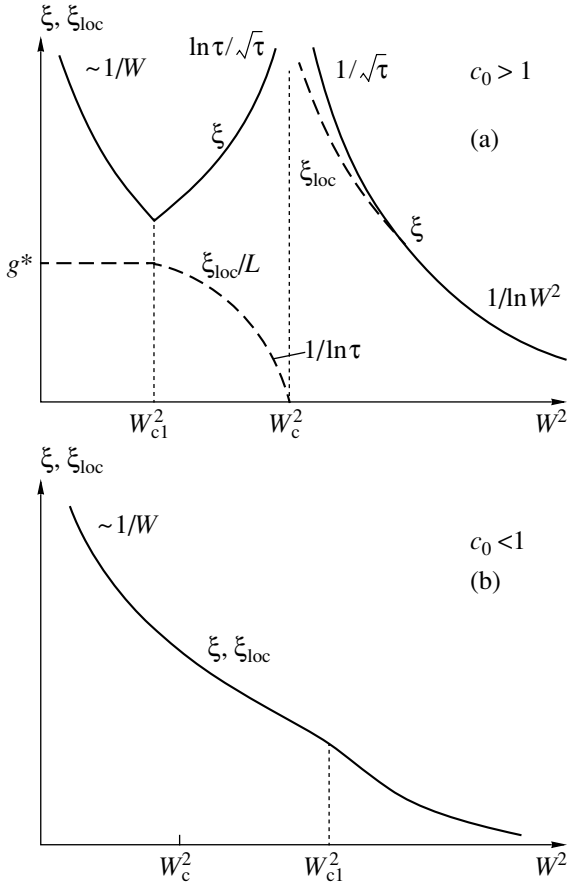

**Fig. 8.** Behavior of characteristic scales for (a) $c_0 > 1$ and (b) $c_0 < 1$.

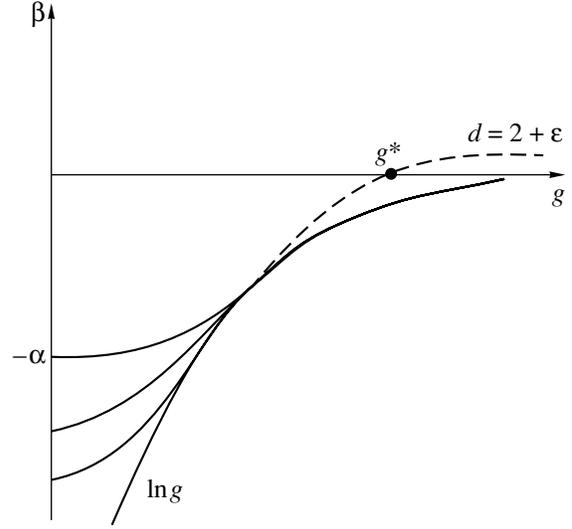

**Fig. 9.** Gell-Mann–Low function $\beta(g)$ is not universal in the small $g$ region.

The latter assumption is not valid in the general case. If power law localization takes place, then $G_L \sim L^{-\alpha}$ and

$$\beta(g) = -\alpha, \quad g \ll 1. \qquad (76)$$

It is clear from the above considerations that the $\beta$-function in the 2D case is not universal for small $g$ and can have a different behavior for different cases (Fig. 9). This conclusion is quite natural from the viewpoint of the general theory of phase transitions [29]. Indeed, scaling is a large-scale property and Eq. (74) has a real sense only for $|\beta(g)| \ll 1$ (that is, in the narrow region near the horizontal axis in Fig. 9), when $g(L)$ slowly changes. In the opposite case, $g(L)$ changes on an atomic scale and there are no grounds for either scale invariance or universality. From the general viewpoint, existence of universal results (75) is rather random and the assumption on universality of $\beta(g)$ for all $g$ [1] is an obvious idealization (see [62] for experimental aspects).

For $d > 2$, Eq. (74) has a fixed point $g^*$ that determines the existence of the Anderson transition. For $d = 2 + \epsilon$, $g^*$ is arranged in the large $g$ region and disappears in the limit $\epsilon \longrightarrow 0$: i.e., the Anderson transition is absent in the 2D case. This conclusion should not be revised, because the true metallic state is indeed absent (Section 3). The transition we have discussed is situated deep in the localized phase and corresponds to switching from one $\beta$-function to another during a change in the external parameters. Consequently, no substantial revision of the weak localization region is necessary.

## 8. CONCLUSIONS

We have shown that the commonly accepted numerical algorithm based on the transfer matrix method is deficient, because the minimal Lyapunov exponent does not obey any scaling. To restore scaling, a modification of the algorithm is necessary which constructively reduces to a change in the number of the Lyapunov exponent in the course of scaling constructions. This modification does not require a significant increase in numerical work, because the higher Lyapunov exponents in any case are determined in the course of evaluating $\gamma_{min}$ [22]. In fact, one can take the old raw data [8–11] and reinterpret them. This will probably resolve the contradictions discussed in Section 1.

Already at this stage one can interpret a strange drift of results for $d = 3$ with increasing system size: $\nu = 0.66$ [7], $\nu = 1.2 \pm 0.3$ [8], $\nu = 1.35 \pm 0.15$ [15], $\nu = 1.54 \pm 0.08$ [16], $\nu = 1.58 \pm 0.02$ [21]. For small $L$, the number of terms in Eq. (16) is not very large and the maximal scale $1/\gamma_{min}$ is indeed related to the correlation length; description of the Anderson transition is rough, but the

674

results are correct in their roughness. For large *L*, the difference between the minimal and effective Lyapunov exponents becomes significant and the results, being formally accurate, become in fact incorrect.

If the concept of the minimal Lyapunov exponent is taken literally, it leads to unambiguous prediction of the 2D phase transition. This transition is of the Kosterlitz–Thouless type and occurs between exponential and power law localization. Modification of the algorithm leads to division of 2D systems into two classes, the first of which is characterized by exponential localization of all states, while in the second class there is a phase transition between exponential and power law localization.

## ACKNOWLEDGMENTS

This work was supported by the Russian Foundation for Basic Research, project 03-02-17519.